# Magnetic-field control of topological electronic response near room temperature in correlated Kagome magnets

Yangmu Li[1,*], Qi Wang[2], Lisa DeBeer-Schmitt[3], Zurab Guguchia[4,5], Ryan D. Desautels[3], Jiaxin Yin[4], Qianheng Du[1,6], Weijun Ren[1,7], Xinguo Zhao[7], Zhidong Zhang[7], Igor A. Zaliznyak[1], Cedomir Petrovic[1,6], Weiguo Yin[1], M. Zahid Hasan[4], Hechang Lei[2,*], John M. Tranquada[1]

*1 Condensed Matter Physics and Materials Science Division, Brookhaven National Laboratory, Upton, New York 11973, USA*
*2 Department of Physics and Beijing Key Laboratory of Opto-electronic Functional Materials & Micro-nano Devices, Renmin University of China, Beijing 100872, China*
*3 Neutron Scattering Division, Oak Ridge National Laboratory, Oak Ridge, Tennessee 37831, USA*
*4 Laboratory for Topological Quantum Matter and Advanced Spectroscopy, Department of Physics, Princeton University, Princeton, NJ 08544, USA.*
*5 Laboratory for Muon Spin Spectroscopy, Paul Scherrer Institute, CH-5232 Villigen PSI, Switzerland*
*6 Materials Science and Chemical Engineering Department, Stony Brook University, Stony Brook, New York 11790, USA*
*7 Shenyang Materials Science National Laboratory, Institute of Metal Research, Chinese Academy of Sciences, Shenyang 110016, China*
*Correspondence to: yangmuli@bnl.gov, hlei@ruc.edu.cn*

Strongly correlated Kagome magnets are promising candidates for achieving controllable topological devices owing to the rich interplay between inherent Dirac fermions and correlation-driven magnetism. Here we report tunable local magnetism and its intriguing control of topological electronic response near room temperature in the Kagome magnet $Fe_3Sn_2$ using small angle neutron scattering, muon spin rotation, and magnetoresistivity measurement techniques. The average bulk spin direction and magnetic domain texture can be tuned effectively by small magnetic fields. Magnetoresistivity, in response, exhibits a measurable degree of anisotropic weak localization behavior, which allows the direct control of Dirac fermions with strong electron correlations. Our work points to a novel platform for manipulating emergent phenomena in strongly-correlated topological materials relevant to future applications.

The tunability of topologically protected states through interactions between magnetism and electronic band structure provides a novel route towards designing complex quantum materials for technological applications. Theoretically, Kagome structures that break time-reversal symmetry have been proposed to host nontrivial topological electronic states with controllability provided by local magnetism [1-5]. Experimental investigations of these proposals have been recently made possible, with the discovery of inherent Dirac/Weyl fermions and magnetism-related Berry curvature in strongly-correlated

Kagome magnets, such as Fe$_3$Sn$_2$, Mn$_3$Sn, and Co$_3$Sn$_2$S$_2$ [6-10]. In addition to being analogues of the graphene lattice, which features massless Dirac bands [11], magnetic interactions in the Kagome magnets lead to exotic magnetic ground states and consequently impact the materials' electronic properties [6,7].

Rhombohedral Fe$_3$Sn$_2$ (space group R$\bar{3}$m) consists of a Fe$_3$Sn bilayer separated by a single Sn layer [Fig. 1(a)] with lattice constants $a = 5.34$ Å and $c = 19.79$ Å in the hexagonal lattice notation. Fe atoms form a breathing Kagome structure that comprises hexagons and equilateral triangles with alternating Fe bond lengths [Fig. 1(b)]. The Fe$_3$Sn bilayer is mainly responsible for both the nontrivial topological states and strongly-correlated magnetism, and thus leads to a direct connection between the two. Electronically, angle-resolved photoemission (ARPES) [6] and scanning tunneling microscopy quasiparticle interference (QPI) measurements [7] revealed massive Dirac bands at low temperatures and transport measurements identified large intrinsic anomalous Hall signals (AHE) from 2 K to 400 K [6,12-15]. Magnetically, Fe$_3$Sn$_2$ is a soft ferromagnet with electron spins residing on Fe atoms [16,17]. The Fe moments are non-collinear and transition from a high-temperature (HT) phase, in which the spins are closer to the $c$ axis, to a low-temperature (LT) phase (below ~ 100 K), in which the spins are close to the Kagome plane [16-18]. Lorentz transmission electron microscopy observes the presence of mesoscopic stripe-like domains of alternating magnetization within the Kagome planes in the HT phase; the domains become much larger, and without clear spin texture, in the LT phase [19-20]. Manipulation of spin texture in the HT phase provides a possible control of the topological band structure near room temperature.

We begin by characterizing the bulk spin texture of the ferromagnetic domains with small-angle neutron scattering (SANS). Neutron scattering probes the magnetic correlations through the interaction between neutron spins and magnetic moments of the sample. The scattering cross-section is proportional to the square of the magnetic moment perpendicular to the neutron momentum transfer, which is in-plane, so that the moment within the Kagome planes contributes selectively to the scattering cross-section while $c$-axis component of the moment always contributes, allowing measurements of magnetic textures. For simplicity, we redefine the $a$-axis to be along a nearest-neighbor Fe-Fe bond, and note that within the Fe Kagome plane there three equivalent axis choices $a$, $a'$, $a''$, separated by 120 degree rotations, as indicated in Fig. 1(b). With respect to the measurement geometry shown in Fig. 1(c), measurements were performed with the selected $a$ axis oriented approximately in the horizontal or vertical direction, and magnetic field was always applied in the horizontal direction.

Representative SANS data for Fe$_3$Sn$_2$ are shown in Fig. 1. The instrument was set such that the magnetic field direction was either approximately parallel [Fig. 1(h)-(l)] or perpendicular to Fe-Fe bond in Kagome plane [Fig. 1(m)-(q)], respectively. In zero magnetic field, the magnetic scattering is nearly isotropic at 100 K [Fig. 1(d)], consistent with the absence of magnetic texture in the LT phase [16,17], and it evolves to a much broader and anisotropic pattern above 100 K [Fig. 1(f), (g); see Supplementary Materials for additional data]. To further examine the magnetic phases, we used muon spin rotation spectroscopy to probe the temperature dependence of the local spontaneous field at the muon stopping site. As shown in Fig. 2(a), the muons sense a sharp change in the local field slightly above 100 K, with a large relaxation rate in the vicinity of the transition. We will focus on behavior at 200 K, which is clearly in the HT phase.

The anisotropic zero-field SANS patterns in Fig. 1(h) and (m) reflect the stripe-like domain texture of the HT phase. The minimum width of the scattering is parallel to the average orientation of the stripe domains, while the large momentum width is in the direction of small domain width. The preferred alignment of stripe spin texture with one of the three $a$ axes can be attributed to shape anisotropy. When the $a$ axis (sample) is rotated by 90°, the anisotropy of the scattering follows. Applying a magnetic field along (approximately) the preferred $a$ axis, the stripe domain orientation first becomes better aligned with the field, and the scattering narrows in the corresponding direction. With increasing field, the magnetic domains with moments along the field direction grow at the expense of the oppositely polarized domains. By 0.3 T, the parallel-magnetization domains dominate, and the scattering collapses to a narrow peak. When the field is applied in-plane but perpendicular to the preferred $a$ axis, as in Fig. 1(o), (p), and (q), the domains rotate. Because the spins prefer to orient along Fe-Fe directions, the reoriented domains occur along the $a'$ and $a''$ axes, whose average direction is close to the field direction. For a quantitative analysis, the scattering cross-sections were fit with a Lattice Lorentzian model [21], for which scattering anisotropy quantifies phase disorder of the mesoscopic magnetic structure.

$$\frac{d\sigma(\boldsymbol{q})}{d\Omega} \sim \sum \frac{\xi_a \xi_\perp}{\pi^2} (1 + q_a^2 \xi_a^2)^{-1} (1 + q_\perp^2 \xi_\perp^2)^{-1} \qquad (1)$$

where $a$ and $\perp$ labels lattice orientations. $\xi$ and $q$ are the corresponding correlation length and momentum transfer (Supplementary Materials). This model assumes a distribution of domains oriented along the $a$ axes, with the number of domains along each direction and the average correlation lengths parallel and perpendicular to the a axes as fitting parameters. Figure 2(b) shows examples of constant intensity contours for each of the 3 domain orientations at two different magnetic fields. The field dependences of the domain populations and correlation lengths are plotted in Fig. 2(c) and (d). The change in the domain distribution as a function of magnetic field, $\Delta n$, for each domain orientation $\phi_M$ depends on the applied field direction $\phi_H$ as $|\sin(\phi_H - \phi_M)|$. The inset of Fig. 2(c) shows that the ratio $\Delta n/|\sin(\phi_H - \phi_M)|$ grows with field in a manner consistent with an effective three-domain (J=1) Brillouin function (Supplementary Materials). The fitted correlation lengths should largely reflect domain size, and we see in Fig. 2(d) that they grow as field increases.

Having identified the field-dependent control of the magnetic domain textures in $Fe_3Sn_2$, we next investigate the electronic response at 200 K (Fig. 3). Resistivity was measured as a function of magnetic field strength and orientation, using the geometry shown in the inset of Fig. 3(d), where the current is applied along an *a* axis. The dominant effect is that the resistivity decreases as the field is applied [Fig. 3(c)]; this can be understood in terms of bulk weak localization for Dirac bands, as we will discuss below. In addition, the resistivity develops an anisotropic response to the orientation of the field relative to the current, as illustrated in Fig. 3(a) and (b). The amplitude of the anisotropic magnetoresistivity (MR) follows the bulk magnetization, as shown in Fig. 3d, while the angle of maximum MR $\phi_{max}$ evolves as field increases. Similar butterfly patterns with an alternating sequence of positive and negative lobes of MR are observed for both S1 and S2 (additional data in Supplementary Materials); differences result from current being aligned to an *a* axis that is or is not magnetically-preferred. These butterfly patterns are different from the angular magnetoresistivity for uniform ferromagnetic materials, for which the resistivity is usually two-fold symmetric and highest when the magnetic field is parallel to the current direction [22]. Moreover, the MR does not vary in a symmetric fashion with respect to the magnetically-preferred *a* axis. At 75 K, when magnetic spin texture is less clear in the LT phase, partial recovery of the magnetoresistivity to the lattice symmetry is observed.

For further analysis, it is convenient to convert resistivity to conductivity, which is obtained by inverting the resistivity tensor. Because the transverse resistivity $\rho_{xy}$ is one order of magnitude smaller than the longitudinal resistivity $\rho_{xx}$ for $Fe_3Sn_2$, conductivity $\sigma_{xx} = \rho_{xx}/(\rho_{xx}^2 + \rho_{xy}^2) \sim 1/\rho_{xx}$. We further define $\sigma_{2D} = \sigma_{xx} \cdot c$, where $c$ is the *c*-axis lattice parameter. The field-induced changes in 2D conductivity for S1 and S2 are presented in Fig. 4(a) and (b), respectively. For each field direction, the conductivity change can be described by weak localization of Dirac bulk bands based on the Hikami-Larkin-Nagaoka equation [23, 24]

$$\Delta\sigma_{2D} \sim -\alpha \frac{e^2}{h} [\frac{1}{2} F\left(\frac{H_\phi + 2H_{SO}}{H}\right) + F\left(\frac{H_\phi + H_{SO}}{H}\right) - \frac{1}{2} F\left(\frac{H_\phi}{H}\right)] \qquad (2)$$

where $F(x) = \psi\left(\frac{1}{2} + x\right) - \ln(x)$, and $\psi$ is the digamma function. $H_\phi$ is the phase-coherence characteristic field and $H_{SO}$ is the spin-orbit characteristic field. The fitting parameters are the prefactor α (expected to be of order 1), $H_\phi$, $H_{SO}$. For S1 and S2, the fitted spin-orbit characteristic lengths $l_{SO} = \sqrt{\hbar/4eH_{SO}}$ are similar and approximately direction independent ($l_{SO}$ ~ 12 nm for S1 and 9 nm for S2). However, $H_\phi$ and α have notable angular dependences with a phase difference of ~ 240° for S1 and S2, corresponding to the difference in orientation of the magnetically-preferred *a* axis. $H_\phi$ features a maximum approximately along *a* axes, which means that the corresponding coherence length, $l_\phi = \sqrt{\hbar/4eH_\phi}$, is a minimum in this direction, with values ranging over 90-170 nm for S1 and 50-170 nm for S2 [Fig. 4(b)]. These numbers are intriguingly consistent with the half magnetic domain size (~ 100-200 nm) measured by SANS at small fields [Fig. 2(d)]. We tentatively compare $H_\phi$ with the magnetic domain distribution and found that the two trace the same angular dependence (~$|\sin(\phi_H - \phi_M)|$), which is indicative of spin texture-determined quasiparticle phase coherence. Furthermore, we observe a modulation of prefactor α that also features an approximate two-fold symmetry [Fig. 4(c)]. The maximum of α does not align with the *a* axes but has a phase shift (~ -20 to -30°). As spin texture and bulk magnetization are much more effectively aligned along *a* axes, this phase shift remains a puzzle. Theoretically, it was proposed for materials with a large Dirac mass, weak localization of bulk bands dominates the electronic response [24], and a variation in Dirac mass impacts α [1,25,26]. The prefactor α extracted from the magnetoresistivity equals a single-band α multiplied by number of bands involved. We estimate the single band α based on previous calculations, without considering band interactions Fig. 4(d) [24]. Within this picture, the results suggest that, similar to α, the Dirac mass is also modulated with a two-fold symmetry, consistent with previous QPI measurements [7]. We illustrate the Dirac band dispersion with varying Dirac masses using modified pseudospin Dirac model [1,27] in Fig. 4(e).

Having demonstrated the effective tuning of local magnetism and its direct control of electronic response in $Fe_3Sn_2$, our results have far reaching implications. Fe spins in $Fe_3Sn_2$ are separated by Sn into bilayer Kagome planes, and their magnetism and topological band structure are quasi-two-dimensional. Interestingly, similar magnetic and electronic properties have recently been discovered for other Kagome systems such as $Co_3Sn_2S_2$ [10] and van der Waals metals such as $Fe_3GeTe_2$ [28,29], for which both low-dimensional ferromagnetism and anomalous Hall effect have been observed. Manipulating the mesoscopic magnetic textures in these materials can give rise to a new control of their topological properties. Enhancing the measurable effects in electronic response tuned by a small applied magnetic field provides a feasible new platform for the realization of functional topological devices.

We gratefully acknowledge helpful discussions with X. Wang, L. Classen, A. Sapkota, G. Cai, and A. M. Tsvelik and thank PSI Bulk µSR Group for invaluable technical support with µSR experiments. Work at Brookhaven National Laboratory is supported by the Office of Basic Energy Sciences, Materials Sciences and Engineering Division, U.S. Department of Energy (DOE) under Contract No. DE-SC0012704. Work at Renmin University of China is supported by the National Key R&D Program of China (Grants No. 2016YFA0300504), the National Natural Science Foundation of China (Grants No. 11574394, 11774423 and 11822412), the Fundamental Research Funds for the Central Universities, and the Research Funds of Renmin University of China (Grants No. 15XNLF06, 15XNLQ07 and 18XNLG14). Work at Shenyang Materials Science National

Laboratory is supported by the National Natural Science Foundation of China (Grant No. 5161192) and the National Key R&D Program of China (Grand No. 2017YFA0206302). The µSR experiment used resources of the low background GPS spectrometer at the πM3 beamline of the Paul Scherrer Institute. The SANS experiment used resources at the High Flux Isotope Reactor, a DOE Office of Science User Facility operated by the Oak Ridge National Laboratory.

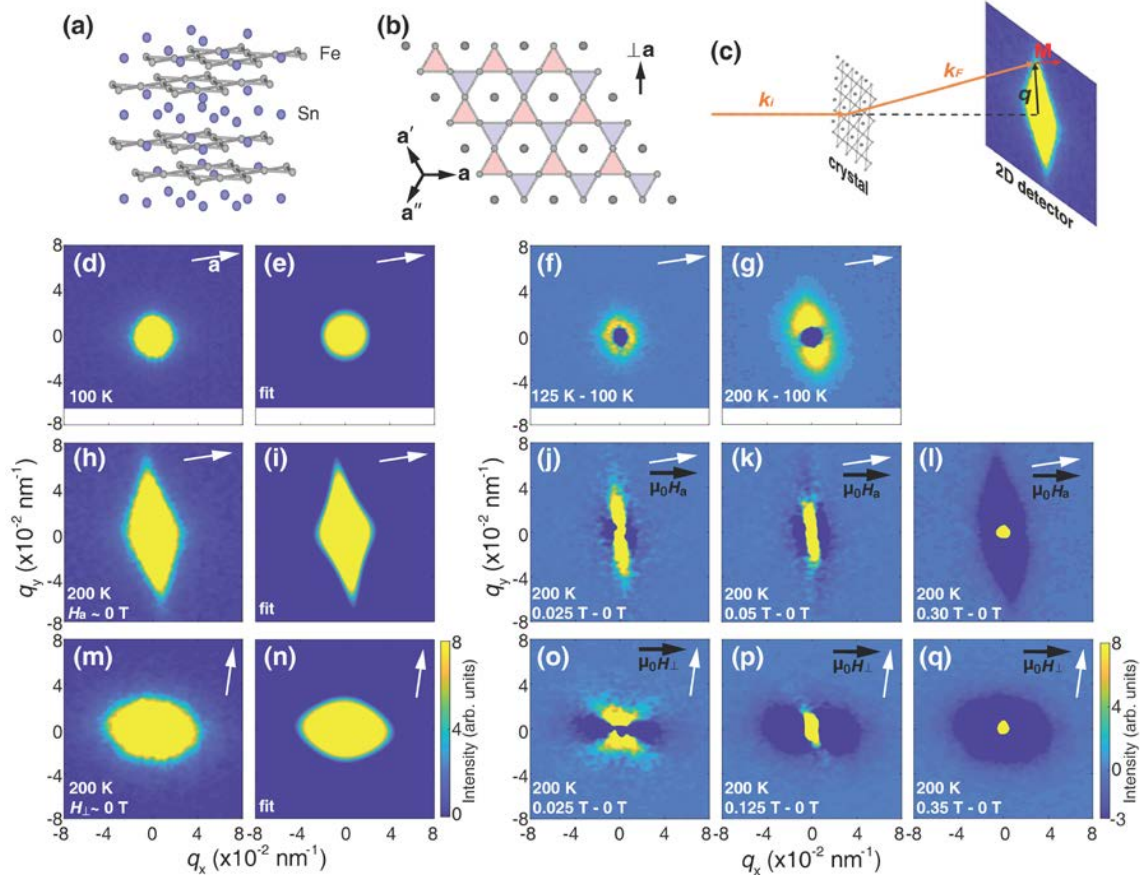

**Fig. 1** $Fe_3Sn_2$ structure and representative SANS data. (a), Structure of $Fe_3Sn_2$. (b), Fe breathing Kagome lattice with short and long Fe bond lengths (red and blue, 2.59 Å and 2.75 Å). $a$, $a'$, $a''$ denote equivalent Fe-Fe bond directions and ⊥ denotes direction perpendicular to $a$. (c), Instrument setup for our SANS measurements. The neutron intensity is proportional to the square of magnetic moment M that perpendicular to neutron kinetic momentum transfer $q$. (d)-(g), SANS scattering patterns in zero magnetic field. (f), (g) show formation of bulk local magnetism when increasing temperature above 100 K. (h)-(l) and (m)-(q) Scattering patterns with magnetic fields parallel and perpendicular to Fe-Fe bond, respectively. (e), (i), (n) are corresponding fits of scattering patterns in d, h, m using the anisotropic Lattice Lorentzian model [22]. White arrows indicate the direction of Fe-Fe bond $a$.

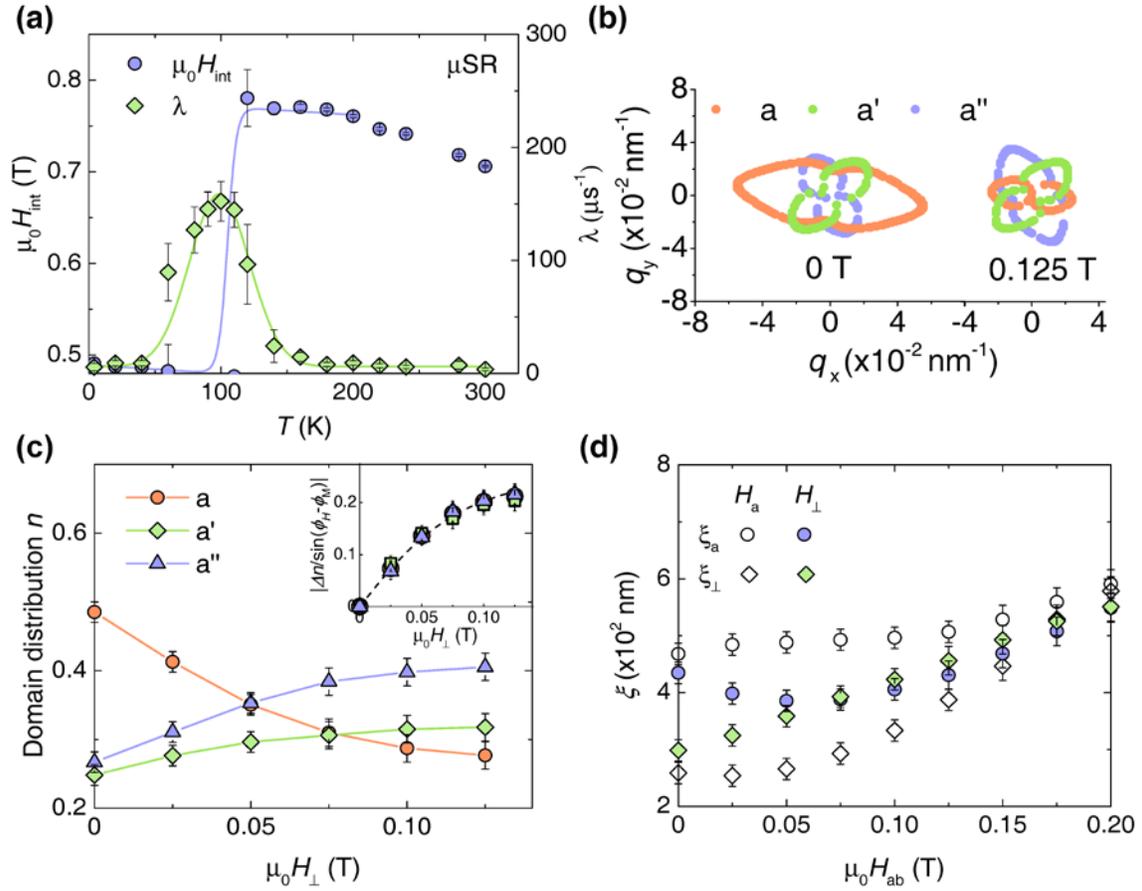

Fig. 2 Magnetic properties. (a), Temperature dependence of the internal magnetic field ($H_{\text{int}}$) and μSR relaxation rate (λ), showing a phase transition around 100 K. (b), Neutron intensity for domain along $a$, $a'$, $a''$ at $H_\perp = 0$ and 0.125 T. (c), Spin direction (magnetic domain) distribution $n$ along $a$, $a'$, $a''$. Insert shows the change of spin distribution $\Delta n = n(H) - n(H=0)$ divided by $|\sin(\phi_H - \phi_M)|$, where $\phi_H$ is the magnetic field direction and $\phi_M$ is the angular direction of a, $a'$, $a''$, respectively. $|\Delta n / \sin(\phi_H - \phi_M)|$ for all spin configuration can be described by an effective three-domain (J=1) Brillouin function. (d), Parallel and perpendicular field dependences of magnetic correlation lengths (domain size). A rearrangement of spin correlation at mesoscopic scale is observed in the case of $H_\perp$.

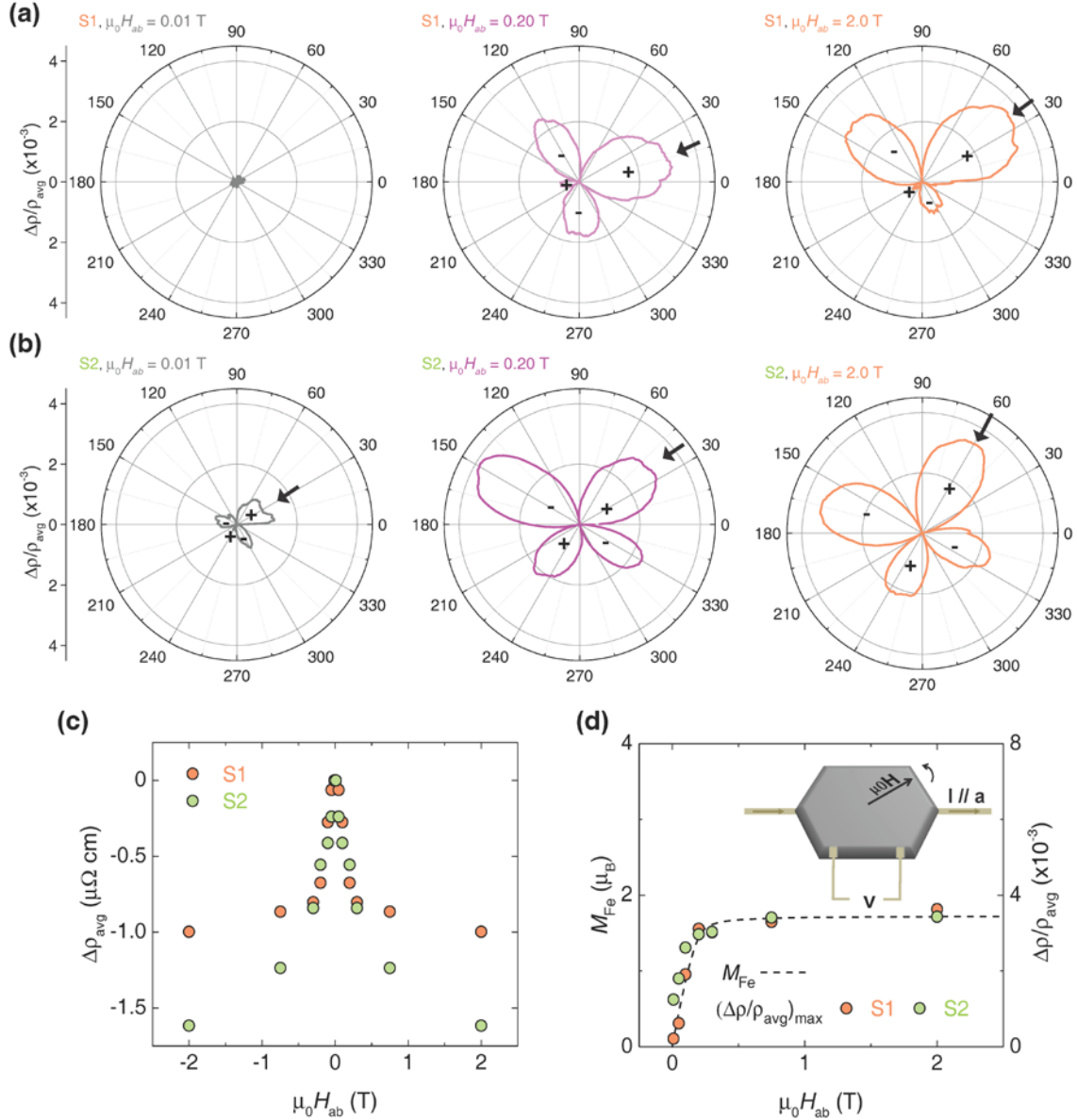

Fig. 3 In-plane magnetoresistivity. (a), (b) Angular variation of in-plane magnetoresistivity at 200 K in different magnetic fields for S1 and S2, respectively. The current is fixed along Fe-Fe bond a (0 degree) and the magnetic field rotates in the plane. The angle between field and current is shown as the polar axis. Magnitude of the butterfly-wing pattern shows the variation of resistivity from its average value $\Delta\rho/\rho_{avg}$, where $\Delta\rho = \rho - \rho_{avg}$. $\rho$ is resistivity at each angle and $\rho_{avg}$ is the averaged resistivity for all field directions. Black arrows indicate $\Delta\rho/\rho_{avg}$ maximum at angle $\phi_{max}$. (c), Magnetic field dependence of $\rho_{avg}$, where $\Delta\rho_{avg} = \rho_{avg}(H) - \rho_{avg}(H = 0)$. (d), Comparison between Fe moment and $\Delta\rho/\rho_{avg}$ maximum as a function of field strength. Inset is an illustration of device geometry.

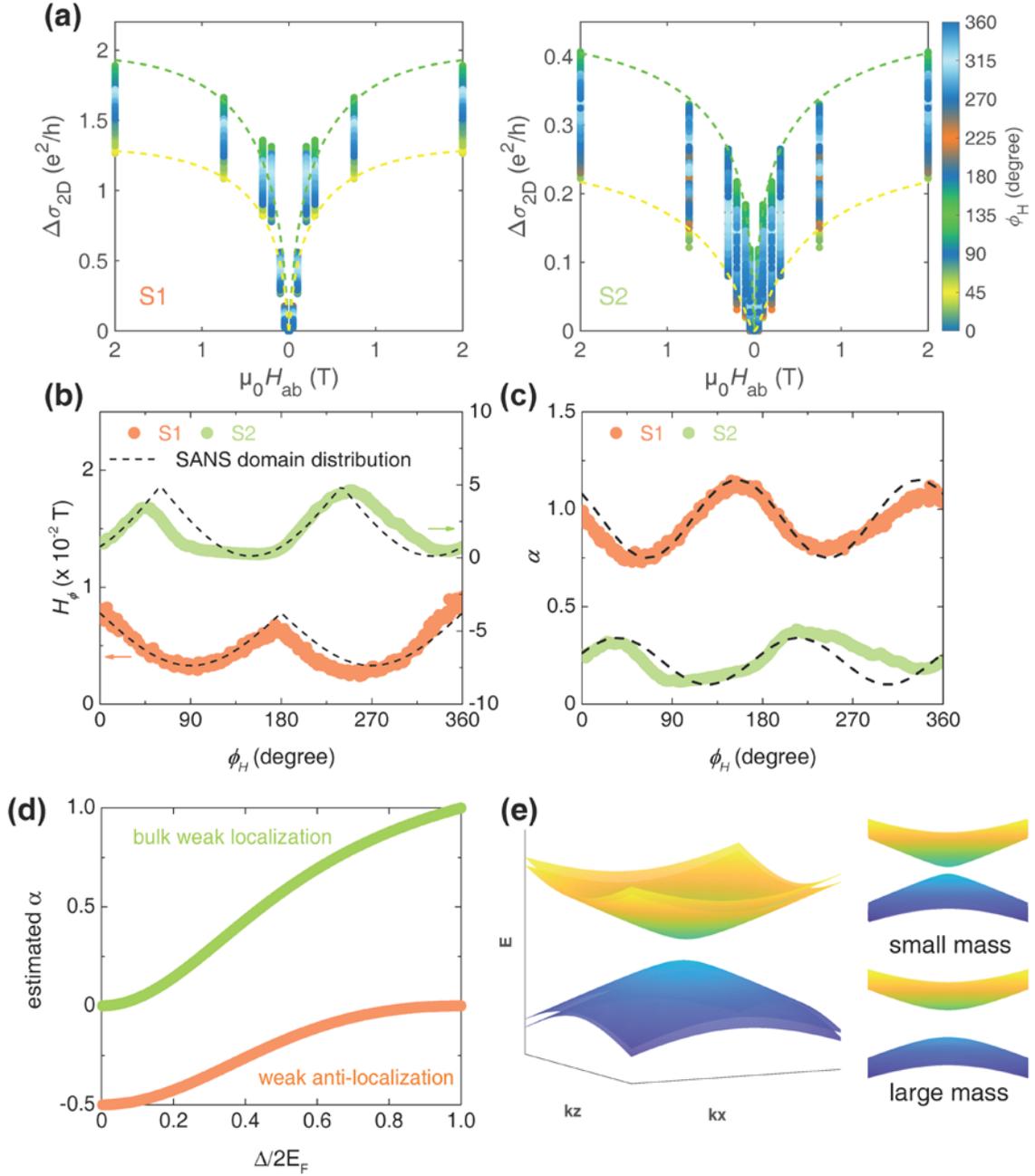

Fig. 4 Weak localization conductivity and Dirac band. (a), Magnetic field dependence of conductivity of all field directions ($\phi_H$ shown as the color bar) for S1 and S2, respectively. $\Delta\sigma_{2D} = \sigma_{2D}(H) - \sigma_{2D}(H=0)$, as explained in the text. Dashed lines present the upper (green) and lower (yellow) boundaries by Hikami-Larkin-Nagaoka equation [23, 24] for weak localization. (b), Phase coherence characteristic field $H_\phi$ and magnetic domain distribution measured by SANS. (c), Prefactor α. (d), Theoretical estimation of single-band α based on Dirac mass [24]. e, Illustration of massive Dirac band structure with the pseudospin Dirac model [1, 25].